\documentclass[preprint,review,12pt]{elsarticle}

\usepackage{graphicx}
\usepackage{amssymb}
 
\usepackage{amsmath}
\usepackage{dsfont}
\usepackage{bm}
\usepackage{eucal}
\usepackage{amsthm}

\usepackage{xcolor}

\def\C{{\mathds C}}

\newcommand{\be}{\begin{equation}}
\newcommand{\ee}{\end{equation}}

\newcommand{\bzero}{{\mbox{\boldmath $0$}}}
\newcommand{\bz}{{\mbox{\boldmath $z$}}}
\newcommand{\bc}{{\mbox{\boldmath $c$}}}
\newcommand{\bee}{{\mbox{\boldmath $e$}}}
\newcommand{\bzeta}{{\mbox{\boldmath $\zeta$}}}
\newcommand{\bgamma}{{\mbox{\boldmath $\gamma$}}}
\newcommand{\bv}{{\mbox{\boldmath $v$}}}

\newcommand{\bu}{{\mbox{\boldmath $u$}}}
\newcommand{\bx}{{\mbox{\boldmath $x$}}}

\newcommand{\bU}{{\mbox{\boldmath $U$}}}
\newcommand{\bA}{{\mbox{\boldmath $A$}}}
\newcommand{\bI}{{\mbox{\boldmath $I$}}}
\newcommand{\bcA}{{\mbox{\boldmath ${\cal A}$}}}
\newcommand{\bB}{{\mbox{\boldmath $B$}}}
\newcommand{\bcB}{{\mbox{\boldmath ${\cal B}$}}}

\newcommand{\bZ}{{\mbox{\boldmath $Z$}}}
\newcommand{\bG}{{\mbox{\boldmath $G$}}}

\newcommand{\bR}{{\mbox{\boldmath $R$}}}
\newcommand{\bM}{{\mbox{\boldmath $M$}}}
\newcommand{\bom}{{\mbox{\boldmath $m$}}}

\newtheorem{theorem}{Theorem}
\newtheorem{lemma}{Lemma}

\newcommand{\test}{\mbox{$
\begin{array}{c}
\stackrel{ \stackrel{\textstyle H_1}{\textstyle >} }{ 
\stackrel{\textstyle <}{ \textstyle  H_0} }

\end{array}
$}}

\def\cC{\mbox{$\CMcal C$}}
\def\cL{\mbox{$\CMcal L$}}
\def\cN{\mbox{$\CMcal N$}}

\begin{document}

\begin{frontmatter}

\title{Adaptive Radar Detection in Heterogeneous Clutter-dominated Environments}

\author[a]{Angelo Coluccia}
\ead{angelo.coluccia@unisalento.it}

\author[b]{Danilo Orlando}
\ead{danilo.orlando@unicusano.it}

\author[a]{Giuseppe Ricci\corref{cor1}}
\ead{giuseppe.ricci@unisalento.it}

\cortext[cor1]{Corresponding author}

\address[a]{Dipartimento di Ingegneria dell'Innovazione, Universit\`a del Salento, Via Monteroni, 73100 Lecce, Italy.}

\address[b]{Engineering Faculty of Universit\`a degli Studi ``Niccol\`o Cusano'', 
via Don Carlo Gnocchi 3, 00166 Roma, Italy}

\begin{abstract}
In this paper, we propose a new solution for the detection problem of a coherent target 
in heterogeneous environments. Specifically, we first assume that clutter returns from different range bins
share the same covariance structure but different power levels. This model meets the experimental evidence
related to non-Gaussian and non-homogeneous scenarios. Then, unlike existing solutions that are based upon
estimate and plug methods, we propose an approximation of the 
generalized likelihood ratio test where the maximizers of the likelihoods are obtained 
through an alternating estimation procedure.
Remarkably, we also prove that such estimation procedure leads to an architecture possessing the 
constant false alarm rate (CFAR) when a specific initialization is used.
The performance analysis, carried out on simulated as well as measured data and in comparison with
suitable well-known competitors, highlights that the proposed architecture can overcome the
CFAR competitors and exhibits a limited loss with respect to the other non-CFAR detectors.
\end{abstract}


\end{frontmatter}

\section{Introduction}
Nowadays, radar systems are ubiquitous in real life with applications ranging from 
military to civil field
\cite{melvin2013principles}. 
One of the main implications of such a rapid 
and endless technological development is that the new operating scenarios have become more challenging.
Moreover, they require
sophisticated signal processing algorithms that were unimaginable a few decades ago due to the
limited computational resources provided by the processing units. 
Such a complexity forces radar engineers to leave aside the classical design assumptions.
In fact, focusing on target detection
algorithms, even though the most common design assumptions, namely, the Gaussian distribution for the clutter and the
homogeneity of data under test and training samples (homogeneous environment)
\cite[and references therein]{Kelly86,robey1992cfar,BOR-Morgan,junBook}, 
allow for a mathematical tractability of the problem,
they are not always valid as corroborated by the experimental 
measurements \cite{Ward,Ward1990,Farina_Gini_Greco_Verrazzani,
Greco_Gini_Rangaswamy,Wardbook}. For instance, in high-resolution radars, especially at low grazing angles, 
clutter is generally modeled as a compound-Gaussian process whose complex envelope results from the product
of a speckle component (obeying the complex Gaussian distribution) and a texture component (that is a real and nonnegative
random process) \cite{Compound1,Compound2}. When observed on sufficiently short 
time intervals, the compound-Gaussian process degenerates into a spherically invariant
random process (SIRP) where the texture can be approximated as a deterministic quantity \cite{Compound1,Compound2}.
In this case, each range bin within the radar reference window is characterized by 
a specific texture value possibly varying over the range.
An asymptotically optimum approximation 
of the generalized likelihood ratio test (GLRT) to detect a coherent signal in the presence 
of interference\footnote{The terms
interference and disturbance are used to denote the joint action of clutter and thermal noise.} 
modeled in terms of a SIRP has been derived in \cite{CLR1995}. Interestingly, such an architecture
coincides with the exact GLRT under the so-called partially-homogeneous environment \cite{kraut1999cfar},
where data under test and training samples 
(are complex Gaussian distributed and) share the same interference covariance
structure but are characterized by their respective 
interference power levels. This model represents an intermediate design step
between the homogeneous and the ``fully-heterogeneous'' environment \cite{9145700}.

Generally speaking, design assumptions that account for possible inhomogeneities in the reference window
are of primary importance in the radar community since training samples may often be contaminated
by power variations over range, clutter discretes, and other outliers. As a consequence,
the volume of homogeneous training data does not allow for reliable estimates (sample-starved scenarios).
Different solutions to this limitation have been conceived in the open literature. For instance, the well-known knowledge-based
paradigm exploits {\em a priori} information at the design stage to reduce the 
requirements in terms of secondary data amount. Other widely used techniques consist of the regularization (or shrinkage)
of the sample covariance matrix towards a given matrix \cite{gerlach1,yuriMGLRT,Tyler,AC}, of clustering training data into
homogeneous subsets \cite{XU2021108127,9321174,9413918}, or of detecting 
and suppressing the outliers \cite{629144,767347,1433140}.

With the above remarks in mind, in this paper, we attack the detection of a coherent target assuming 
the fully-heterogeneous scenario, where data vectors obey the complex Gaussian distribution and 
share the same structure of the interference covariance matrix
but different power levels. As stated above, this model meets the experimental evidence related to 
non-Gaussian, clutter-dominated environments. Unlike \cite{Conte-DeMaio-Ricci1,Loss1,Greco-Gini,Conte-Demaio,Pascal2008},
where estimate and plug solutions have been conceived and assessed, herein, we devise a suitable approximation 
of the GLRT. Specifically, such an approximation is dictated by the fact that the straightforward application of
the maximum likelihood approach for parameter estimation under each hypothesis leads to intractable mathematics
(at least to the best of authors' knowledge).
Therefore, we resort to a cyclic optimization procedure that, at each iteration,
moves towards a local stationary point of the likelihoods \cite{Stoica_alternating}. 
Remarkably, we prove that the newly proposed architecture 
can exhibit the costant false alarm rate (CFAR)
property with respect to the clutter covariance structure or the power levels or both according 
to the specific seeds for the alternating procedure.

Finally, the performance assessment is conducted over simulated as well as real recorded data and in
comparison with estimate and plug solutions. 
The latter are grounded on the normalized matched filter (NMF) \cite{CLR1995}
coupled with the normalized \cite{Eusipco94}, recursive \cite{Loss1}, and persymmetric \cite{Conte-Demaio} estimates
of the clutter covariance matrix.


The remainder of the paper is organized as follows: the next section is devoted to 
the design of the detector for heterogeneous clutter-dominated environments.
Section \ref{Sec:performance} assesses its performance 
also in comparison with the aforementioned counterparts. Moreover, it provides
two propositions that establish the CFAR behavior of the proposed architecture.
Finally, Section \ref{Sec:conclusions} contains some concluding remarks and draws future research lines.

\subsection{Notation}
Vectors and matrices are denoted by boldface lower-case and upper-case letters, respectively.
Symbols $\det(\cdot)$, $(\cdot)^T$, $(\cdot)^\dag$, $(\cdot)^{-1}$, and
$(\cdot)^-$ denote the determinant, transpose, 
conjugate transpose, inverse, and generalized inverse, respectively. 
As to numerical sets, 
$\C$ is the set of 
complex numbers, $\C^{N\times M}$ is the Euclidean space of $(N\times M)$-dimensional 
complex matrices, and $\C^{N}$ is the Euclidean space of $N$-dimensional 
complex vectors. 
We denote by $\bee_1$ the first vector of the canonical basis for $\C^{N}$;
$|z|$ and $\overline{z}$ denote the modulus and the complex conjugate of the complex number $z$, respectively.
The identity matrix of size $N\times N$ is indicate by $\bI_N$.
The acronym RV means random variable.
The acronym PDF stands for probability density function and
we write $\bx\sim\cC\cN_N(\bom, \bM)$ if $\bx$ is a 
complex normal $N$-dimensional random vector with mean value $\bom$ and positive definite covariance matrix $\bM$.

\section{Clutter-dominated Environment: GLRT-based Design}

Let us assume a system that collects $N$ (space, time, or space-time) samples 
from the range cell under test (CUT).
The problem of detecting the possible presence of a coherent return in the given CUT can be  formulated as the following
hypothesis testing problem:
\begin{equation}
\left\{
\begin{array}{ll}
H_{0}: & \bz \sim \cC \cN_N (\bzero, \bR), \\ 
 & \bz_{k} \sim \cC\cN_N (\bzero, \gamma_k \bR), \quad k=1, \ldots, K,\\
H_{1}: & \bz \sim \cC\cN_N (\alpha \bv, \bR), \\ 
 & \bz_{k} \sim {\cC\cN}_N (\bzero, \gamma_k \bR), \quad k=1, \ldots, K,
\end{array} 
\right.
\label{FO-HT}
\end{equation}
where 
$\bz \in \C^{N}$ is the vector of samples from the CUT,
$\bv \in \C^{N}$ is the known (space, time, or space-time) steering vector,
$\alpha \in \C$ is an unknown parameter accounting for channel propagation, radar cross section of the target, etc.,
$\bR \in \C^{N \times N}$ is an unknown positive definite
covariance matrix accounting for the 
common correlation among samples of the disturbance,  
the $\bz_k$s are the $N$-dimensional secondary data vectors,
and $\gamma_1, \ldots, \gamma_K>0$ are
unknown parameters taking into account the different power levels of secondary 
data with respect to the CUT  (without loss of generality).
Finally, we suppose that $K \geq N$.
\medskip

We want to determine the GLRT for the problem at  hand; it is given by
$$
\frac{\max_{\alpha, \bR,  \bgamma} f_1( \bz, \bZ; \alpha, \bR,  \bgamma)
}{
\max_{\bR, \bgamma} f_0( \bz, \bZ; \bR,  \bgamma)
} \test \eta
$$
where $\bgamma=[\gamma_1 \cdots \gamma_K]^T$, $\bZ=[\bz_1 \cdots \bz_K]$,  and
$\eta$ is the threshold to be set according to the desired probability of false alarm ($P_{fa}$)
while $f_1$ and $f_0$ denote the
joint PDFs of 
the CUT and secondary data under $H_1$ and $H_0$, respectively. We have that
the PDF is given by
\begin{align}
\nonumber
&f_1( \bz, \bZ; \alpha, \bR,  \bgamma) =
\frac{1}{\pi^{N (K+1)}} \frac{1}{ \prod_{k=1}^K \gamma_k^{N}}
\frac{1}{\det^{K+1}(\bR)} 
\\ &\times \exp\left\{
-  \left((\bz-\alpha \bv)^{\dag}\bR^{-1}(\bz-\alpha \bv)  +
\sum_{k=1}^K \frac{\bz_k^{\dag}\bR^{-1}\bz_k}{\gamma_k}
\right)
\right\}
\label{PDF_H1}
\end{align}
under $H_1$ and under $H_0$ by
\begin{align*}
&f_0( \bz, \bZ; \bR,  \bgamma) =
\frac{1}{\pi^{N (K+1)}} \frac{1}{\prod_{k=1}^K \gamma_k^{N}}
\frac{1}{\det^{K+1}(\bR)} 
\\ &\times \exp\left\{-    \left(\bz^{\dag}\bR^{-1}\bz +
\sum_{k=1}^K \frac{\bz_k^{\dag}\bR^{-1}\bz_k}{\gamma_k}
\right)
\right\}.
\end{align*}

The maximizers of the PDFs with respect to $\bR$, given the remaining parameters,
are given by
$$
\widehat{\bR}_1= \frac{1}{K+1}
\left[ \left(\bz-\alpha \bv \right)\left(\bz-\alpha \bv \right)^{\dag} + \sum_{k=1}^K \frac{\bz_k\bz_k^{\dag}}{\gamma_k}
\right]
$$
and
$$
\widehat{\bR}_0= \frac{1}{K+1}
\left[ \bz\bz^{\dag} + \sum_{k=1}^K \frac{\bz_k\bz_k^{\dag}}{\gamma_k}
\right]
$$
under $H_1$ and $H_0$, respectively.
The corresponding partially-compressed likelihoods can be written as 
\begin{align}
\nonumber
&l_1( \alpha, \widehat{\bR}_1,   \bgamma;
\bz, \bZ) =
\left( \frac{K+1}{e\pi} \right)^{N (K+1)} \frac{1}{ \prod_{k=1}^K \gamma_k^{N}}
\\ &\times
\frac{1}{\det^{K+1}\left[ \left(\bz-\alpha \bv \right)\left(\bz-\alpha \bv \right)^{\dag} + \sum_{k=1}^K \frac{\bz_k\bz_k^{\dag}}{\gamma_k}
\right]} 
\label{PDF_H1_1}
\end{align}
under $H_1$ and under $H_0$ by
\begin{align}
\nonumber
&l_0(  \widehat{\bR}_0,   \bgamma; \bz, \bZ) =
\left( \frac{K+1}{e\pi} \right)^{N (K+1)} \frac{1}{ \prod_{k=1}^K \gamma_k^{N}}
\\ &\times
\frac{1}{\det^{K+1}\left[ \bz\bz^{\dag} + \sum_{k=1}^K \frac{\bz_k\bz_k^{\dag}}{\gamma_k}
\right]}.
\label{PDF_H0_1}
\end{align}

At authors' knowledge maximization with respect to both
$\alpha$ and the $\gamma_k$s under $H_1$ and with respect to the $\gamma_k$s under $H_0$ cannot be conducted in closed form. For this reason, we consider an alternating procedure \cite{Stoica_alternating}. More precisely, under $H_1$, given the $t$th estimate of the $\gamma_k$s, 
say $\widehat{\gamma}_k^{1,(t)}$, $k=1, \ldots, K$,
we can maximize eq. (\ref{PDF_H1_1}) with respect to
$\alpha$, thus obtaining $\widehat{\alpha}^{(t+1)}$; 
moreover, given $\widehat{\alpha}^{(t+1)}$ and the $\widehat{\gamma}_k^{1,(t+1)}$s, $k=1, \ldots, h-1$,  
together with the $\widehat{\gamma}_k^{1,(t)}$, $k=h+1, \ldots, K$, 
we can maximize eq. (\ref{PDF_H1_1}) with respect to
$\gamma_h$, thus obtaining $\widehat{\gamma}_h^{1,(t+1)}$.
Under $H_0$ we can proceed in a similar way to maximize
eq. (\ref{PDF_H0_1}) with respect to the 
$\gamma_k$s, $k=1, \ldots, K$, assuming that $\alpha=0$.

In order to maximize eq. (\ref{PDF_H1_1}) with respect to $\alpha$ 
we minimize the following function
$$
f_1(\alpha)=
{\det}\left[ \left(\bz-\alpha \bv \right)\left(\bz-\alpha \bv \right)^{\dag} + \bA^{(t)}
\right]
$$
where $\bA^{(t)}=\sum_{k=1}^K \frac{\bz_k\bz_k^{\dag}}{\widehat{\gamma}_k^{1,(t)}}$ is a positive definite matrix. It turns out that
$$
f_1(\alpha)={\det} \left(\bA^{(t)} \right) \
\left[1+ \left(\bz-\alpha \bv \right)^{\dag} \left(\bA^{(t)}\right)^{-1}
\left(\bz-\alpha \bv \right)
\right]
$$
is minimized by
\be
\widehat{\alpha}^{(t+1)}=
\frac{\bv^{\dag} \left(\bA^{(t)}\right)^{-1} \bz}
{\bv^{\dag} \left(\bA^{(t)}\right)^{-1} \bv }.
\label{eq:alpha_estimate}
\ee
Let us introduce the function
\begin{eqnarray*}
f_2(\gamma_h) &=&
\gamma_h^{N} \
{\det}^{K+1}\left[ \left(\bz-\widehat{\alpha}^{(t+1)} \bv \right)\left(\bz-\widehat{\alpha}^{(t+1)} \bv \right)^{\dag} 
\right. \\
&+& \left.
\sum_{k=1}^{h-1} \frac{\bz_k\bz_k^{\dag}}{\widehat{\gamma}_k^{1,(t+1)}}
+\sum_{k=h+1}^K \frac{\bz_k\bz_k^{\dag}}{\widehat{\gamma}_k^{1,(t)}}
+\frac{1}{\gamma_h}\bz_h\bz_h^{\dag}
\right]
\\ &=&
\gamma_h^{N} \
{\det}^{K+1}\left[ \bB^{1,(t+1)}_h
+\frac{1}{\gamma_h}\bz_h\bz_h^{\dag}
\right]
\\ &=&
\gamma_h^{N} \ {\det}^{K+1}\left( \bB^{1,(t+1)}_h \right)
\\ &\times& \left( 1
+\frac{1}{\gamma_h}\bz_h^{\dag}\left( \bB^{1,(t+1)}_h \right)^{-1} \bz_h
\right)^{K+1}
\end{eqnarray*}
where
\begin{eqnarray}
\nonumber
\bB^{1,(t+1)}_h
&=&
\left(\bz-\widehat{\alpha}^{(t+1)} \bv \right)\left(\bz-\widehat{\alpha}^{(t+1)} \bv \right)^{\dag} + \sum_{k=1}^{h-1} \frac{\bz_k\bz_k^{\dag}}{\widehat{\gamma}_k^{1,(t+1)}}
\\ &+& \sum_{k=h+1}^K \frac{\bz_k\bz_k^{\dag}}{\widehat{\gamma}_k^{1,(t)}}
\label{eq:B_under_H1}
\end{eqnarray}
is a positive definite matrix.
Since $\lim_{\gamma_h \rightarrow 0}
f_2(\gamma_h)=+\infty$ and $\lim_{\gamma_h \rightarrow +\infty}
f_2(\gamma_h)=+\infty$, it follows that
the minimum is attained at a stationary point of $f_2$.
Moreover, the derivative of $f_2$ is given by
\begin{eqnarray*}
\frac{d}{d \gamma_h} f_2(\gamma_h) &=&
N \gamma_h^{N-1} \ {\det}^{K+1}\left( \bB^{1,(t+1)}_h \right)
\left( 1
+\frac{1}{\gamma_h}\bz_h^{\dag}\left( \bB^{1,(t+1)}_h \right)^{-1} \bz_h
\right)^{K+1}
\\ &+&
\gamma_h^{N} \ {\det}^{K+1}\left( \bB^{1,(t+1)}_h \right)
\ (K+1) 
\left( 1+\frac{1}{\gamma_h}\bz_h^{\dag}\left( \bB^{1,(t+1)}_h \right)^{-1} \bz_h
\right)^{K} 
\\ &\times& \left( - \frac{1}{\gamma_h^2}\bz_h^{\dag}\left( \bB^{1,(t+1)}_h \right)^{-1} \bz_h \right).
\end{eqnarray*}
Thus, the minimizer is given by
\be
\widehat{\gamma}_h^{1,(t+1)}=\frac{K+1-N}{N} \ \bz_h^{\dag}\left( \bB^{1,(t+1)}_h \right)^{-1} \bz_h.
\label{eq:gamma_h_estimate}
\ee

Summarizing, to implement the proposed detector, we start with an initial estimate
of the $\gamma_k$s under $H_1$, say $\widehat{\gamma}_k^{1,(0)}$s,
and use eqs. (\ref{eq:alpha_estimate}) and (\ref{eq:gamma_h_estimate}) to obtain $\widehat{\alpha}^{(1)}$
and $\widehat{\gamma}_k^{1,(1)}$, $k=1, \ldots, K,$ respectively, 
and after $t_{\max}$ iterations  (see the stopping criterion below)
$\widehat{\alpha}^{(t_{\max})}$
and $\widehat{\gamma}_k^{1,(t_{\max})}$, $k=1, \ldots, K.$
Similarly, we compute the estimate of the $\gamma_k$s under $H_0$,
say $\widehat{\gamma}_k^{0,(t_{\max})}$, $k=1, \ldots, K.$
As a matter of fact,
we have to modify eq. (\ref{eq:gamma_h_estimate}) by replacing $\bB^{1,(t+1)}_h$ of eq. (\ref{eq:B_under_H1})
with
\be
\bB^{0,(t+1)}_h=
\bz\bz^{\dag} + \sum_{k=1}^{h-1} \frac{\bz_k\bz_k^{\dag}}{\widehat{\gamma}_k^{0,(t+1)}}
+\sum_{k=h+1}^K \frac{\bz_k\bz_k^{\dag}}{\widehat{\gamma}_k^{0,(t)}}.
\label{eq:B_under_H0}
\ee

The entire procedure may terminate after $t_{\max}$ iterations, where $t_{\max}$ is such that
\be
\Delta\cL(t_{\max})=\frac{|\cL(t_{\max})-\cL(t_{\max}-1)|}{|\cL(t_{\max}-1)|}<\epsilon,
\ee
where $\epsilon>0$ and $\cL(t)$ is the log-likelihood function at the $t$th iteration, 
or $t_{\max}$ is the maximum allowable number of iterations set according 
to the selected compromise between performance and computational requirements.

Finally, we obtain the following approximation of the  GLRT statistic
\be
\frac{
\left[
\frac{
\prod_{k=1}^K \widehat{\gamma}_k^{0,(t_{\max})}}
{\prod_{k=1}^K \widehat{\gamma}_k^{1,(t_{\max})}}
\right]^{\frac{N}{K+1}}
\det\left[ \bz \bz^{\dag} + \sum_{k=1}^K \frac{\bz_k\bz_k^{\dag}}{\widehat{\gamma}_k^{0,(t_{\max})}}
\right]
}
{\det\left[ \left(\bz-\widehat{\alpha}^{(t_{\max})} \bv \right)\left(\bz-\widehat{\alpha}^{(t_{\max})} \bv \right)^{\dag} + \sum_{k=1}^K \frac{\bz_k\bz_k^{\dag}}{\widehat{\gamma}_k^{1,(t_{\max})}}
\right]}.
\label{eq:GLRT_approximation}
\ee

\section{Performance assessment}
\label{Sec:performance}
This section is devoted to the analysis of the proposed detector and consists of two subsections. 
The first subsection
proves that the proposed detector can be CFAR with respect to 
both the matrix $\bR$ and the 
parameters 
$\gamma_1, \ldots, \gamma_K$ under the design assumptions.
The aim of the latter subsection is twofold. Firstly, it assesses 
the performance of the detector in comparison to natural competitors and also in the presence of  
possible mismatches between the nominal and the actual operating conditions by resorting to Monte Carlo simulation.
In particular, we investigate the effects due to the presence of a small, but non-negligible, 
thermal noise component  (actually we will assume non-Gaussian, clutter-dominated environments). 
Finally, this subsection contains also illustrative examples obtained using real recorded data.

\subsection{Theoretical analysis: CFAR property}\label{sec:theor}

In the following we will prove that the proposed detector can possess the CFAR property with respect to
$\bR$ and the $\gamma_k$s under the design assumptions.
We start with the following preliminary result.

\begin{lemma}
Suppose that the joint distribution of the entries of the random vectors
$
\widehat{\bgamma}^{1,(t)}=[\widehat{\gamma}_1^{1,(t)}, \cdots, \widehat{\gamma}_K^{1,(t)}]^T,
$
$
\widehat{\bgamma}^{0,(t)}=[\widehat{\gamma}_1^{0,(t)}, \cdots, \widehat{\gamma}_K^{0,(t)}]^T
$ 
and $\bU \bR^{-1/2} [\bz \ \bZ]$, with
$\bU$ a unitary matrix
such that $\bU \bR^{-1/2} \bv$ is aligned with  $\bee_1$ (the first vector of the canonical basis),
is independent of $\bR$ under $H_0$. Then, the joint distribution of the entries of the random vectors
$
\widehat{\bgamma}^{1,(t+1)}$, 
$\widehat{\bgamma}^{1,(t)}$, 
$\widehat{\bgamma}^{0,(t+1)}$,
$\widehat{\bgamma}^{0,(t)}$, $\bU \bR^{-1/2} [\bz \ \bZ]
$
is also independent of $\bR$ under $H_0$. 
\end{lemma}

\begin{proof}
See Appendix A.
\end{proof}

We are now in the condition to prove the CFAR property 
of the proposed detector with respect to $\bR$.
In fact, the following theorem holds true.

\begin{theorem}
Suppose that
the joint distribution of the RVs $\widehat{\gamma}_k^{1,(0)}$, 
$\widehat{\gamma}_k^{0,(0)}$, $k=1, \ldots, K,$ and
the entries of $\bU \bR^{-1/2} [\bz \ \bZ]$, is independent of $\bR$ under $H_0$ (see Lemma 1 for the definition of $\bU$).
Then, the distribution of the statistic (\ref{eq:GLRT_approximation})   is independent of $\bR$ under the $H_0$ hypothesis (and provided that the design assumptions are satisfied).
\end{theorem}

\begin{proof}
See Appendix A.
\end{proof}

Another preliminary result is

\begin{lemma}
Suppose that, under $H_0$,
$\widehat{\gamma}_k^{1,(t)}=\gamma_k g^{1,(t)}_k(\bz, \bG)$ and $\widehat{\gamma}_k^{0,(t)}=\gamma_k g^{0,(t)}_k(\bz, \bG)$, $k=1, \ldots, K$,
with $\bG=[\bz_1/\sqrt{\gamma_1} \cdots \bz_K/\sqrt{\gamma_K}]$.
It turns out that 
$\widehat{\gamma}_k^{1,(t+1)}=\gamma_k g^{1,(t+1)}_k(\bz, \bG)$, $\widehat{\gamma}_k^{0,(t+1)}=\gamma_k g^{0,(t+1)}_k(\bz, \bG)$, $k=1, \ldots, K$,
and, in addition, $\widehat{\alpha}^{(t+1)}=h^{(t+1)}(\bz, \bG)$  
is independent of $\gamma_1, \ldots, \gamma_K$.
\end{lemma}

\begin{proof}
See Appendix A.
\end{proof}

Now, we can prove the CFAR property with respect to the scale factors through the following theorem.

\begin{theorem}
Suppose that, under $H_0$,  $\widehat{\gamma}_k^{1,(0)}$ and $\widehat{\gamma}_k^{0,(0)}$, $k=1, \ldots, K$,
satisfy the assumptions of Lemma 2 for 
$\widehat{\gamma}_k^{1,(t)}$ and $\widehat{\gamma}_k^{0,(t)}$, $k=1, \ldots, K$. Then
the decision statistic  (\ref{eq:GLRT_approximation}) possesses the CFAR property with respect to the $\gamma_k$s.
\end{theorem}

\begin{proof}
See Appendix A.
\end{proof}

As final remark of this subsection, we give an example of initialization of
the $\gamma_k$s
that satisfies the assumptions of Theorem 1 and 2 and, hence, guarantees the CFAR property of the proposed detector.
The initialization is 
$$
\widehat{\gamma}_k^{i,(0)}=
\bz_k^{\dag} (\bz \bz^{\dag})^-  \bz_k, \quad k=1, \ldots, K, \ i=0,1,
$$
where we recall that $(\cdot)^-$ denotes a generalized inverse of the matrix argument. 
As a matter of fact, 
it is straightforward to check that 
the assumptions of Theorem 2 are satisfied.
Moreover, we have that
\begin{eqnarray*}
\bz_k^{\dag} (\bz \bz^{\dag})^-  \bz_k
&=&
\left( \bU \bR^{-1/2} \bz_k \right)^{\dag} 
\bU \bR^{1/2}
(\bz \bz^{\dag})^-  \bR^{1/2} \bU^{\dag} \bU \bR^{-1/2}\bz_k
\\ &=&
\left( \bU \bR^{-1/2} \bz_k \right)^{\dag} 
\left( \bU \bR^{-1/2}
(\bz \bz^{\dag}) \bR^{-1/2} \bU^{\dag} \right)^- \bU \bR^{-1/2}\bz_k
\\ &=&
\left( \bU \bR^{-1/2} \bz_k \right)^{\dag} 
\left( \bU \bR^{-1/2} \bz
\left( \bU \bR^{-1/2} \bz\right)^{\dag} \right)^- \bU \bR^{-1/2}\bz_k
\end{eqnarray*}
where we have used property (9) in 3.6.1 of \cite{Handbookofmatrices}. Thus, also the assumptions of Theorem 1 are satisfied. 

\subsection{Detection performance: simulated and real recorded data}
In what follows, we analyze the performance of the proposed detector
against natural competitors. In particular, we consider the NMF detector \cite{CLR1995}, coupled with 
three different estimates of the disturbance covariance matrix:
\begin{itemize}
\item the so-called normalized sample covariance matrix defined in \cite{Eusipco94};
\item the estimate relying on the recursive procedure devised in \cite{Loss1};
\item the recursive estimate exploiting the persymmetric structure of the covariance matrix proposed in \cite{Conte-Demaio}.
\end{itemize}
The considered competitors are reference benchmarks for detection in heterogeneous scenarios.
The number of iterations for the above recursive estimators is set to three, since this number is sufficient to 
guarantee an acceptable convergence as corroborated by the related literature.

Starting the analysis from the simulated data, 
a desired $P_{fa} = 10^{-3}$ is assumed and the performance is assessed by Monte Carlo simulation 
with $100/P_{fa}$ independent trials to set the thresholds as well as to estimate the $P_{fa}$;
the $P_d$ values are estimated based upon $10^{3}$
trials. Moreover, we set $N=8$, $K=16$, and use a temporal steering vector with zero Doppler. 
The non-homogeneous disturbance is then generated according to the compound-Gaussian model, with the texture 
distributed as the square root of a Gamma  random variable with 
parameters $(\nu,1/\nu)$ (so that the mean square value is unitary) and the complex normal speckle 
with exponentially-shaped covariance matrix, i.e., the $(m_1,m_2)$th entry of 
the matrix $\bR$ is given by $ \rho^{|m_1-m_2|}$, $\rho=0.95$.
Additional white (thermal) noise is also considered in some numerical examples.
Specifically, as first step, we show the performance results under the design assumptions, 
i.e., without thermal noise, then we assess the performance under mismatched conditions, that is considering 
the presence of both clutter and thermal noise. 
To this end, we adopt the following general definition for the signal-to-noise ratio (SNR)
\begin{equation}
\mathrm{SNR} = |\alpha|^2 \bv^\dag (\bR +\sigma^2 \bm{I}_N)^{-1} \bv,
\label{eq:SNR}
\end{equation}
where $\sigma^2>0$ is the thermal noise power and can be set to zero when the
proposed detector is assessed under the design assumption.

The number of iterations for the proposed algorithm is set to $20$.
\begin{figure}
\centering
\includegraphics[width=9cm]{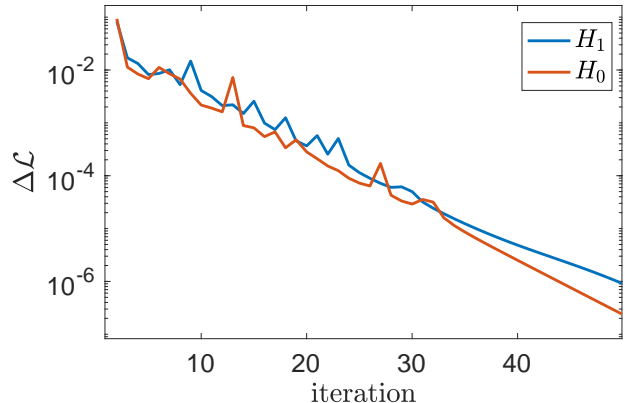}
\caption{Average log-likelihood relative variation between consecutive iterations for the proposed detector (both hypotheses).}
\label{fig1}
\end{figure}
This value is justified by Fig. \ref{fig1}
where  we plot the averaged (over $10^5$ Monte Carlo trials) log-likelihood relative variation under both hypotheses
as a function of the number of iterations. It turns out that such a variation 
quickly decreases with the iteration number and  is no greater than $10^{-3}$ after 20 
iterations.

\begin{figure}
\centering
\includegraphics[width=9cm]{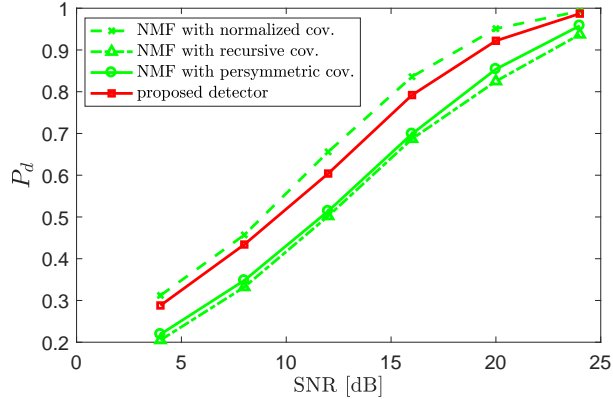}
\caption{$P_d$ vs SNR of the proposed detector, 
in comparison with natural competitors, in the clutter-only case.}
\label{fig2}
\end{figure}

\begin{figure}
\centering
\includegraphics[width=9cm]{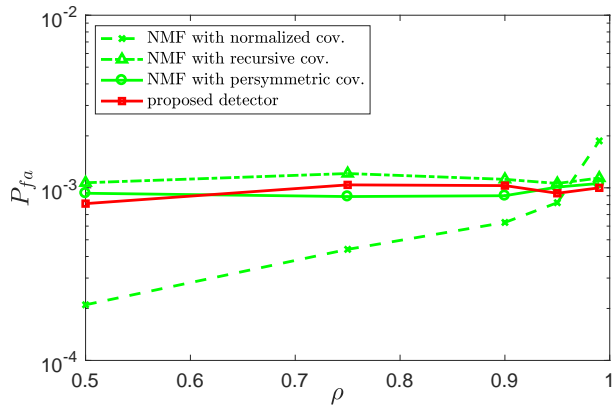}
\caption{Sensitivity analysis of $P_{fa}$ with respect to mismatched one-lag correlation coefficient of the clutter covariance matrix (nominal value $\rho=0.95$), in the clutter-only case.}
\label{fig3}
\end{figure}

\begin{figure}
\centering
\includegraphics[width=9cm]{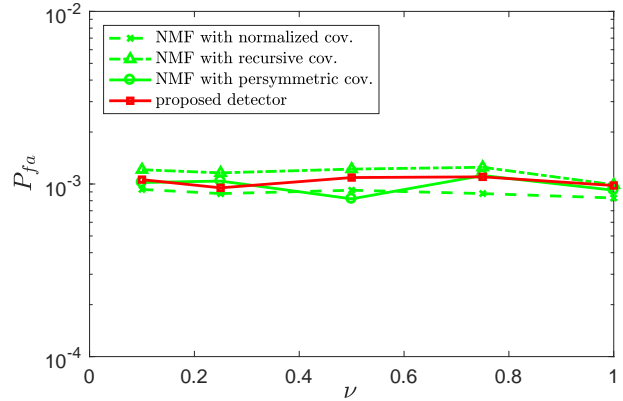}
\caption{Sensitivity analysis of $P_{fa}$  with respect to mismatched texture parameter of the clutter (nominal value $\nu=0.5$), in the clutter-only case.}
\label{fig4}
\end{figure}

\begin{figure}
\centering
\includegraphics[width=9cm]{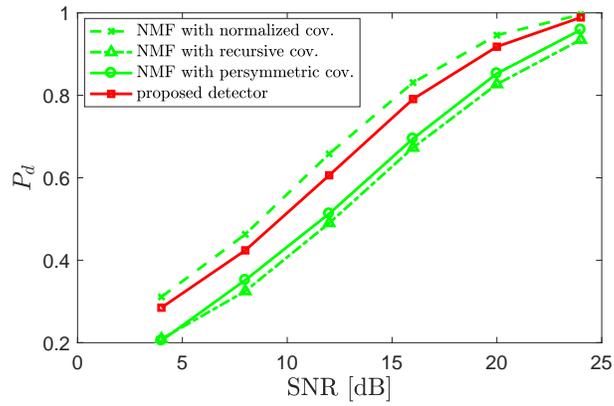}
\caption{$P_d$ vs SNR of the proposed detector, in comparison with natural competitors, in the clutter plus thermal noise case.}
\label{fig5}
\end{figure}

Fig. \ref{fig2} reports the curves of $P_d$ versus SNR under 
the design assumptions (i.e., assuming $\sigma^2=0$) 
and setting $\nu=0.5$ (this value is also used in Fig. \ref{fig5}).
The curves show that the proposed 
architecture achieves better performance than the considered competitors except for the
NMF coupled with the normalized sample covariance matrix. However, 
the advantage of the latter is basically a consequence of 
its sensitivity 
to mismatches of the clutter distribution, i.e., it lacks the CFAR property. 
This fact is confirmed by Fig. \ref{fig3} that 
contains the $P_{fa}$ curves for the considered decision schemes
as functions of the actual one-lag correlation coefficient. It is clearly visible that only 
the NMF with normalized sample covariance matrix is sensitive to mismatches with respect to $\rho$. 
In particular, the corresponding $P_{fa}$ values become unacceptably higher in the upper range, whereas the 
other detectors are theoretically CFAR with respect to $\rho$ in the absence of thermal noise.
From the analysis in Sec. \ref{sec:theor}, we also know that all the detectors are CFAR with respect to $\nu$, 
and this is confirmed in the numerical results shown 
in Fig. \ref{fig4} (again computed over $10^5$ Monte Carlo trials).

\begin{figure}
\centering
\includegraphics[width=9cm]{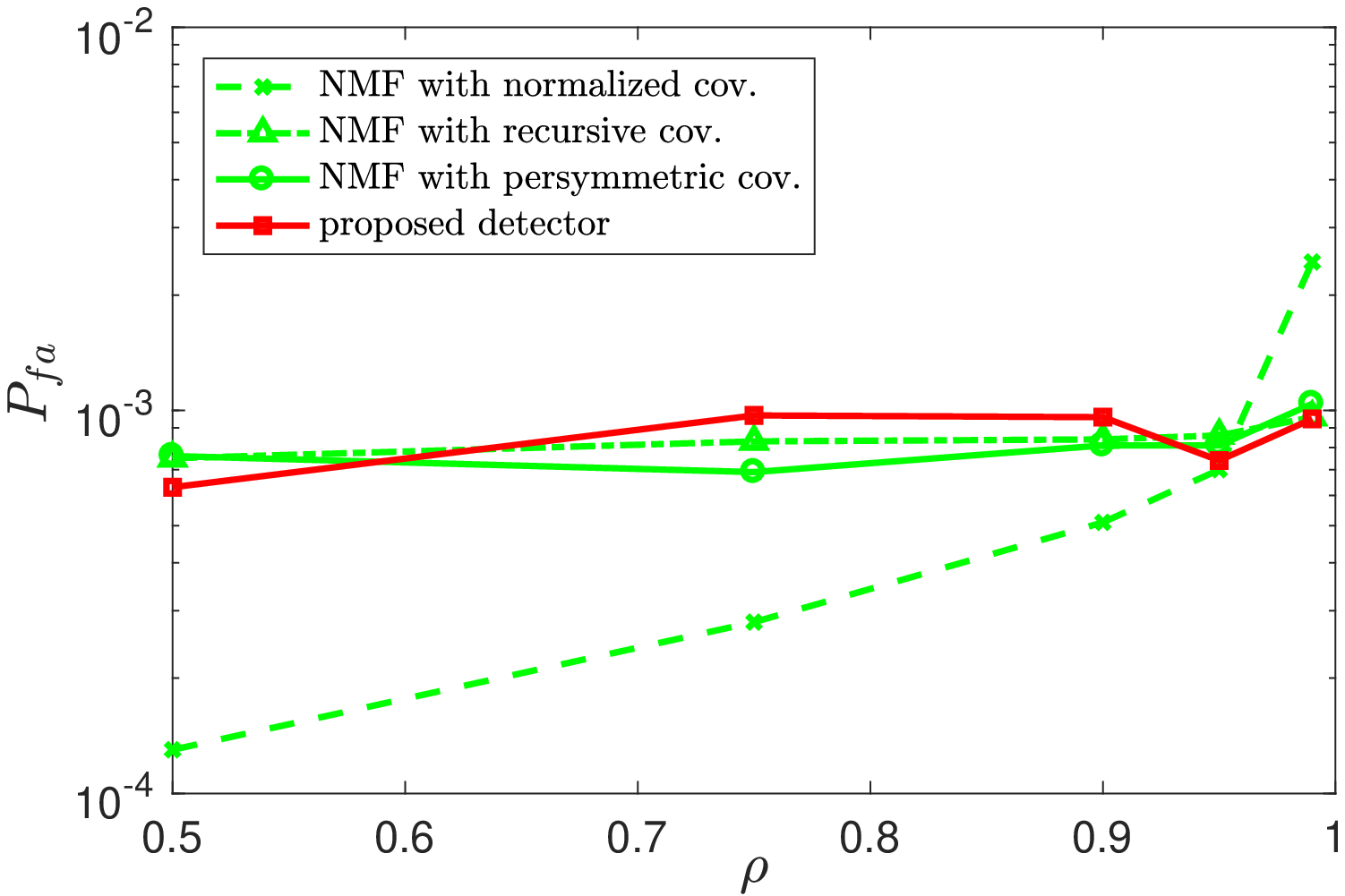}
\caption{Sensitivity analysis of $P_{fa}$ with respect to mismatched one-lag correlation coefficient of the clutter covariance matrix (nominal value $\rho=0.95$), in the clutter plus thermal noise case.}
\label{fig6}
\end{figure}

\begin{figure}
\centering
\includegraphics[width=9cm]{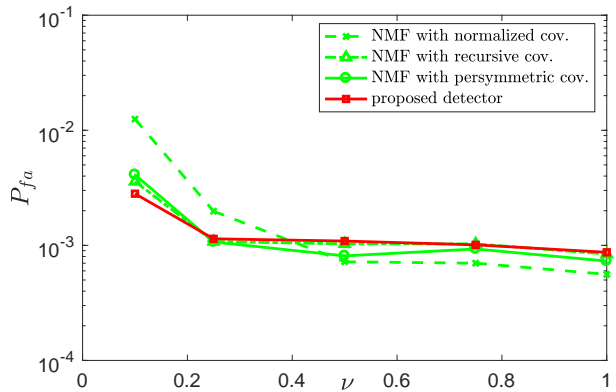}
\caption{Sensitivity analysis of $P_{fa}$  with respect to mismatched texture parameter of the clutter (nominal value $\nu=0.5$), in the clutter plus thermal noise case.}
\label{fig7}
\end{figure}

We now consider the analysis in the 
presence of thermal noise (i.e., $\sigma^2>0$) deviating from the design assumptions. 
The clutter-to-noise ratio is $40$ dB and we consider $\nu=0.5$.
Fig. \ref{fig5} shows that the proposed decision rule still  overcomes the other CFAR competitors
in the whole range of SNR. 
As for the NMF coupled with
the normalized sample covariance matrix, it is slightly superior to the proposed detector, but
its $P_{fa}$ is more sensitive to clutter parameter variations as shown in 
Figs. \ref{fig6} and \ref{fig7}.
Indeed, Fig. \ref{fig6} shows that, while the presence of thermal noise 
hampers strict CFARness for all detectors, only the NMF with normalized 
covariance matrix experiences a nonnegligible sensitivity to mismatches on $\rho$.
In Fig. \ref{fig7}, we show the curves of $P_{fa}$ as a function of the actual texture parameter. 
The results indicate that  CFAR detectors  share almost the same weak sensitivity with respect to $\nu$
but for the NMF coupled with the normalized covariance matrix, which experiences a more marked increase in $P_{fa}$ for small values of $\nu$. 

\begin{figure}
\centering
\includegraphics[width=9cm]{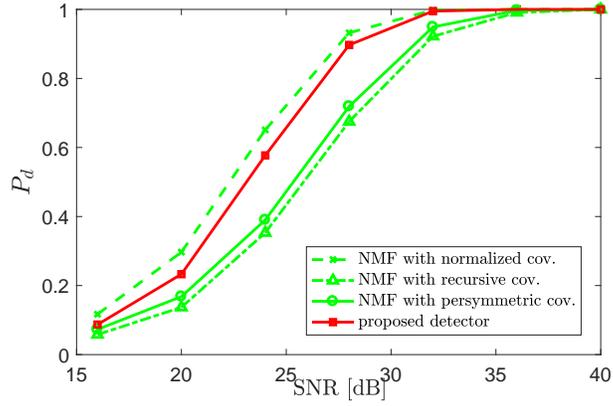}
\caption{$P_d$ vs SNR of the proposed detector, in comparison with natural competitors, on the Phase One real dataset (range bin 30 for both threshold setting and $P_d$ evaluation).}
\label{fig8}
\end{figure}

\begin{figure}
\centering
\includegraphics[width=9cm]{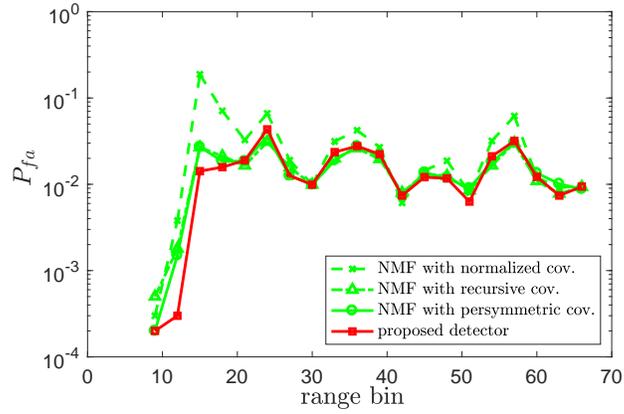}
\caption{$P_{fa}$ sensitivity for the different range bins, with threshold computed on range bin 30, on the Phase One real dataset.}
\label{fig9}
\end{figure}

In the second and final part of this subsection,
we assess the performance of the considered architectures using real L-band 
land clutter data, recorded in 1985 using the MIT Lincoln Laboratory Phase One 
radar at the Katahdin Hill site, MIT Lincoln Laboratory. We consider the dataset contained 
in the file ``H067038.3'', which is composed of
30720 temporal returns from 76 range cells with HH polarization. 
More details can be
found in \cite{Billingsley01,937467} and references therein. Fig. \ref{fig8} 
reports the $P_d$ for range bin $30$, as function of the SNR defined 
by\footnote{Notice that the covariance matrix used in the SNR definition
might be different from the actual covariance matrix.}
\eqref{eq:SNR}. The threshold is set 
by considering $8$-dimensional temporal vectors with $5$ pulses of overlap, 
so as to obtain a sufficient number of snapshots to match the $100/P_{fa}$ rule, for $P_{fa}= 10^{-2}$.
Thus, the detectors work at the same $P_{fa}$. The target is then added synthetically as done 
for the simulated data. It turns out that the proposed architecture confirms its excellent behavior
with detection performance very close to that 
of NMF coupled with normalized covariance matrix, which however is not CFAR. 
Indeed, in order to investigate the CFAR behavior of the considered detectors, in 
Fig. \ref{fig9},  we report the estimates of the $P_{fa}$ on different range bins, using the same values 
of the threshold obtained for range bin 30. The proposed detector mostly guarantees
$P_{fa}$ values lower than or equal to the nominal $P_{fa}$, whereas 
the NMF coupled with the normalized covariance matrix exhibits more significant 
deviations towards larger values of $P_{fa}$.

\begin{figure}
\centering
\includegraphics[width=9cm]{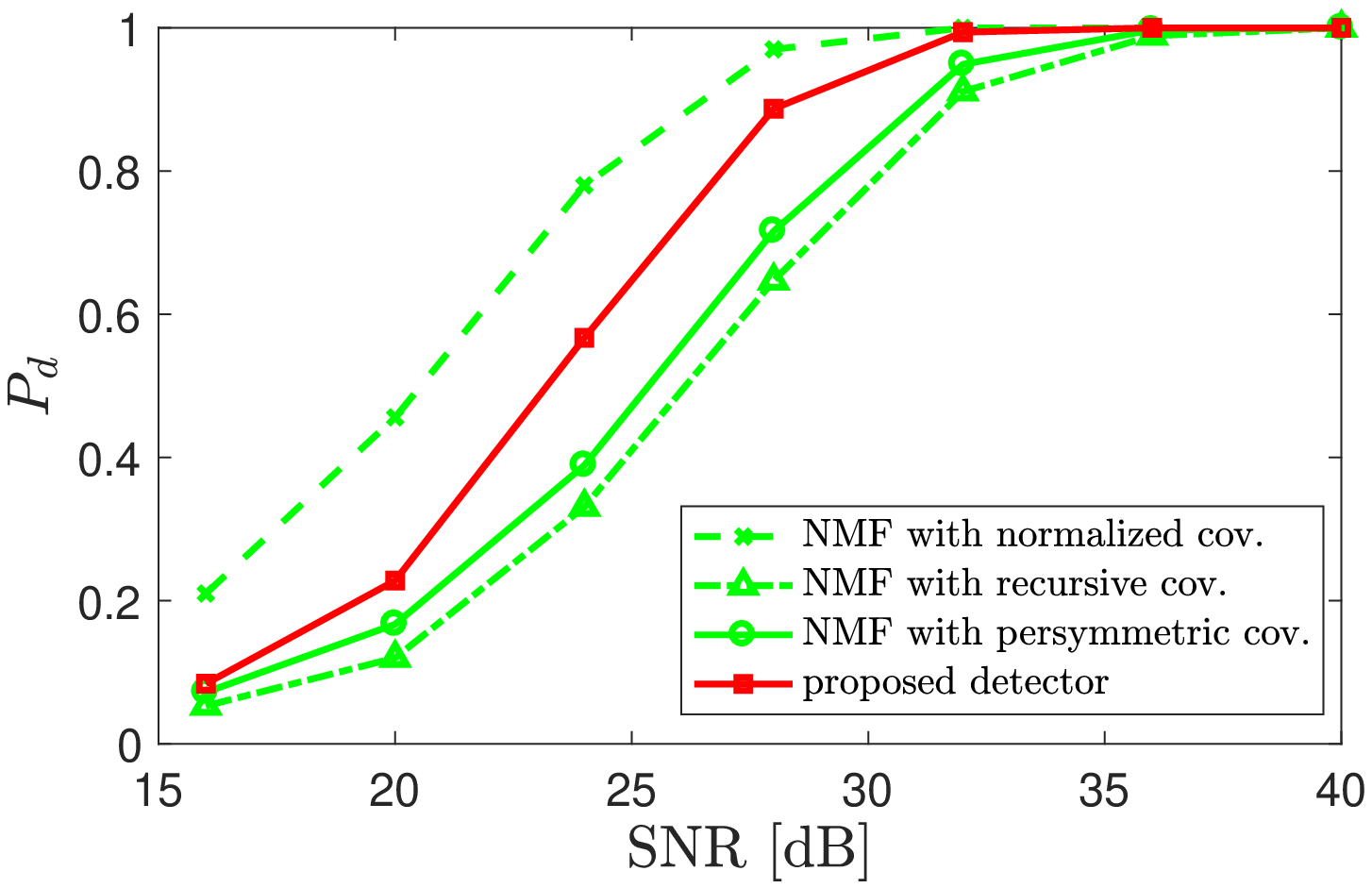}
\caption{$P_d$ vs SNR of the proposed detector, in comparison with natural competitors, on the Phase One real dataset (white noise for  threshold setting and range bin 30 for $P_d$ evaluation).}
\label{fig10}
\end{figure}

\begin{figure}
\centering
\includegraphics[width=9cm]{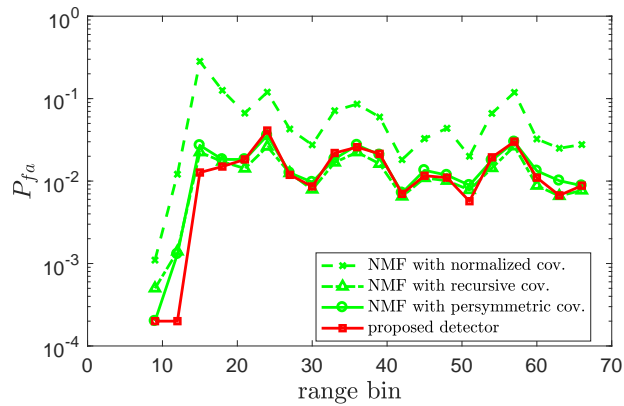}
\caption{$P_{fa}$ sensitivity for the different range bins, with threshold computed on range bin 30, on the Phase One real dataset (white noise for  threshold setting).}
\label{fig11}
\end{figure}

We conclude the performance analysis by showing in Figs. \ref{fig10} and \ref{fig11} 
the results on the Phase One data when 
the threshold is synthetically set on white noise, i.e., $\bR=\bm{I}_N$ and $\sigma^2=0$,
while $P_d$ and $P_{fa}$ are evaluated on real data. Results confirm the goodness of the proposed approach.


\section{Conclusion}
\label{Sec:conclusions}
We have derived an approximation to the  GLRT to detect a coherent target 
in heterogeneous environments. The considered scenario includes clutter returns from different range bins
that share the same covariance structure but different power levels as 
experimentally measured in real environments where the Gaussian assumption is no longer valid.
To solve this problem, we have conceived an alternating estimation procedure 
allowing for an approximation of the 
GLRT unlike existing solutions that are based upon
estimate and plug methods. At the analysis stage,
we have performed a theoretical and experimental investigation using both synthetic and real data.
Remarkably, we have proved that the proposed estimation procedure leads to an architecture possessing the 
CFAR property when a specific initialization is used.
The illustrative examples have shown that the proposed solution 
represents an excellent compromise between detection performance and CFAR behavior.
As a matter of fact, it provides better detection performance than the CFAR competitors
and exhibits a limited loss with respect to the non CFAR detector obtained by coupling
the NMF with the normalized sample covariance matrix.
Finally, the design of architectures accounting
for the different components of the interference covariance matrix is currently 
under investigation and represents a promising new research line.

\section*{Appendix A}
\renewcommand{\theequation}{A.\arabic{equation}}

\section*{Proof of Lemma 1}

First observe that
$\widehat{\alpha}^{(t+1)}$ can be rewritten as 
\be
\widehat{\alpha}^{(t+1)}=
\frac{\bu^{\dag} \left(\bcA^{(t)}\right)^{-1} \bzeta}
{\bu^{\dag} \left(\bcA^{(t)}\right)^{-1} \bu }
\label{eq:alpha_estimate_whitened}
\ee
where
$\bzeta=\bR^{-1/2} \bz$, $\bu=\bR^{-1/2} \bv$, and
$
\bcA^{(t)}=\sum_{k=1}^K \frac{\bzeta_k\bzeta_k^{\dag}}{\widehat{\gamma}_k^{1,(t)}}
$
with
$\bzeta_k=\bR^{-1/2} \bz_k.$ Notice that $\bzeta$ and the $\bzeta_k$s are independent random vectors with a Gaussian distribution that is independent of $\bR$ under $H_0$.
Moreover, we can rotate $\bu$ in order to obtain a vector with 
a nonzero first entry and the remaining entries equal to zero, i.e.,
$\bU \bu=[u \ 0 \cdots 0]^T$ by using a proper unitary matrix 
$\bU$ that does not modify the statistical characterization of  $\bzeta$ and the $\bzeta_k$s. Thus, we can also write
\be
\widehat{\alpha}^{(t+1)}=
 \frac{\overline{u} \left(\bc^{(t)}\right)^{ \dag} \bU \bzeta}
{|u|^2 \left(\bc^{(t)}\right)^{ \dag} \bee_1 }
\label{eq:alpha_estimate_whitened_and_rotated}
\ee
with $\left(\bc^{(t)}\right)^{ \dag}$ the first row of the inverse of the matrix 
$\bU \bcA^{(t)} \bU^{\dag}$ (recall that
$\bee_1$ is the first vector of the canonical basis for $\C^{N \times 1}$). Moreover, $\widehat{\gamma}_1^{1,(t+1)}$, given by eq. (\ref{eq:gamma_h_estimate}), can be rewritten as
$$
\widehat{\gamma}_1^{1,(t+1)}=\frac{K+1-N}{N} \ \bzeta_1^{\dag}\left( \bcB^{1,(t+1)}_1 \right)^{-1} \bzeta_1,
$$
with
\begin{eqnarray*}
\nonumber
\bcB^{1,(t+1)}_1
&=&
\left(\bzeta-\widehat{\alpha}^{(t+1)} \bu \right)\left(\bzeta-\widehat{\alpha}^{(t+1)} \bu \right)^{\dag} + \sum_{k=2}^K \frac{\bzeta_k\bzeta_k^{\dag}}{\widehat{\gamma}_k^{1,(t)}},
\end{eqnarray*}
and also as
\be
\widehat{\gamma}_1^{1,(t+1)}=\frac{K+1-N}{N} \ \left(\bU \bzeta_1 \right)^{\dag}\left( \bU\bcB^{1,(t+1)}_1 \bU^{\dag} \right)^{-1} \bU \bzeta_1
\label{eq:gamma_h_estimate_transformed}
\ee
with
\begin{align*}
\bU\bcB^{1,(t+1)}_1 \bU^{\dag} &=
\left(\bU \bzeta- \frac{\left(\bc^{(t)}\right)^{ \dag} \bU \bzeta}
{\left(\bc^{(t)}\right)^{ \dag} \bee_1 } \bee_1 \right)\left(\bU \bzeta-\frac{\left(\bc^{(t)}\right)^{ \dag} \bU \bzeta}
{\left(\bc^{(t)}\right)^{ \dag} \bee_1 } \bee_1 \right)^{\dag} 
\\ &+ 
\sum_{k=2}^K \frac{\bU \bzeta_k \left( \bU \bzeta_k \right)^{\dag}}{\widehat{\gamma}_k^{1,(t)}}
\end{align*}
where we have used eq. (\ref{eq:alpha_estimate_whitened_and_rotated}).
It is apparent that, since the joint distribution of
$\widehat{\gamma}_1^{1,(t)},
\ldots, \widehat{\gamma}_K^{1,(t)}
$ 
and
the entries of $\bU \bR^{-1/2} [\bz \ \bZ]$
is independent of $\bR$ under $H_0$, also
the joint distribution of 
$ \widehat{\gamma}_1^{1,(t+1)}, \widehat{\gamma}_1^{1,(t)},
\ldots, \widehat{\gamma}_K^{1,(t)}
$
and
the entries of $\bU \bR^{-1/2} [\bz \ \bZ]$
is independent of $\bR$ under $H_0$. Similarly,
$\widehat{\gamma}_2^{1,(t+1)}$, given by eq. (\ref{eq:gamma_h_estimate}), can be rewritten as
$$
\widehat{\gamma}_2^{1,(t+1)}=\frac{K+1-N}{N} \ \bzeta_2^{\dag}\left( \bcB^{1,(t+1)}_2 \right)^{-1} \bzeta_2,
$$
with
\begin{eqnarray*}
\nonumber
\bcB^{1,(t+1)}_2
&=&
\left(\bzeta-\widehat{\alpha}^{(t+1)} \bu \right)\left(\bzeta-\widehat{\alpha}^{(t+1)} \bu \right)^{\dag} + \frac{\bzeta_1\bzeta_1^{\dag}}{\widehat{\gamma}_1^{1,(t+1)}}+
\sum_{k=3}^K \frac{\bzeta_k\bzeta_k^{\dag}}{\widehat{\gamma}_k^{1,(t)}},
\end{eqnarray*}
and also as
$$
\widehat{\gamma}_2^{1,(t+1)}=\frac{K+1-N}{N} \ \left(\bU \bzeta_2 \right)^{\dag}\left( \bU\bcB^{1,(t+1)}_2 \bU^{\dag} \right)^{-1} \bU \bzeta_2
$$
with
\begin{align*}
\bU\bcB^{1,(t+1)}_2 \bU^{\dag} &=
\left(\bU \bzeta- \frac{\left(\bc^{(t)}\right)^{ \dag} \bU \bzeta}
{\left(\bc^{(t)}\right)^{ \dag} \bee_1 } \bee_1 \right)\left(\bU \bzeta-\frac{\left(\bc^{(t)}\right)^{ \dag} \bU \bzeta}
{\left(\bc^{(t)}\right)^{ \dag} \bee_1 } \bee_1 \right)^{\dag} 
\\ &+ 
\frac{\bU \bzeta_1 \left( \bU \bzeta_1 \right)^{\dag}}{\widehat{\gamma}_1^{1,(t+1)}}
+
\sum_{k=3}^K \frac{\bU \bzeta_k \left( \bU \bzeta_k \right)^{\dag}}{\widehat{\gamma}_k^{1,(t)}}.
\end{align*}
Thus, the joint distribution of
$ \widehat{\gamma}_1^{1,(t+1)}, 
\widehat{\gamma}_2^{1,(t+1)}, \widehat{\gamma}_1^{1,(t)},
\ldots, \widehat{\gamma}_K^{1,(t)}
$
and
the entries of $\bU \bR^{-1/2} [\bz \ \bZ]$
is independent of $\bR$ under $H_0$.
Iterating this reasoning it is also straightforward to show that the joint distribution of
$
\widehat{\bgamma}^{1,(t+1)}, 
\widehat{\bgamma}^{1,(t)}
$
and
the entries of $\bU \bR^{-1/2} [\bz \ \bZ]$
and eventually
that of
$
\widehat{\bgamma}^{1,(t+1)}, 
\widehat{\bgamma}^{1,(t)}, 
\widehat{\bgamma}^{0,(t+1)},
\widehat{\bgamma}^{0,(t)}
$
and
the entries of $\bU \bR^{-1/2} [\bz \ \bZ]$
is independent of $\bR$.

\section*{Proof of Theorem 1}

Iterative application of Lemma 1, starting with the assumption that the joint distribution of the RVs $\widehat{\gamma}_k^{1,(0)}$, 
$\widehat{\gamma}_k^{0,(0)}$, $k=1, \ldots, K,$
and
the entries of $\bU \bR^{-1/2} [\bz \ \bZ]$ is independent of $\bR$,
leads to a joint distribution of the RVs
$
\widehat{\bgamma}^{1,(t_{\max})}, 
\widehat{\bgamma}^{1,(t_{\max}-1)}, 
\widehat{\bgamma}^{0,(t_{\max})},
\widehat{\bgamma}^{0,(t_{\max}-1)}
$
and
the entries of $\bU \bR^{-1/2} [\bz \ \bZ]$
independent of $\bR$.
It follows that
the statistic of the proposed detector, given by eq. (\ref{eq:GLRT_approximation}), can be written as
$$
\frac{
\left[
\frac{
\prod_{k=1}^K \widehat{\gamma}_k^{0,(t_{\max})}}
{\prod_{k=1}^K \widehat{\gamma}_k^{1,(t_{\max})}}
\right]^{\frac{N}{K+1}}
\det\left[ \bU \bzeta \bzeta^{\dag} \bU^{\dag}+ \sum_{k=1}^K \frac{\bU\bzeta_k\bzeta_k^{\dag}\bU^{\dag}}{\widehat{\gamma}_k^{0,(t_{\max})}}
\right]
}
{\det\left[ \left(\bU \bzeta- \frac{\left(\bc^{(t_{\max}-1)}\right)^{ \dag} \bU \bzeta}
{\left(\bc^{(t_{\max}-1)}\right)^{ \dag} \bee_1 } \bee_1 \right)\left(\bU \bzeta-\frac{\left(\bc^{(t_{\max}-1)}\right)^{ \dag} \bU \bzeta}
{\left(\bc^{(t_{\max}-1)}\right)^{ \dag} \bee_1 } \bee_1 \right)^{\dag}+ \sum_{k=1}^K \frac{\bU\bzeta_k\bzeta_k^{\dag}\bU^{\dag}}{\widehat{\gamma}_k^{1,(t_{\max})}}
\right]},
$$
where $\left(\bc^{(t_{\max}-1)}\right)^{ \dag}$,
defined after eq. (\ref{eq:alpha_estimate_whitened_and_rotated}) in the proof of Lemma 1, is a function of $\widehat{\bgamma}^{1,(t_{\max}-1)}$
and
the entries of $\bU \bR^{-1/2} \bZ$
and hence, the theorem is proved.

\section*{Proof of Lemma 2}

First observe that $\widehat{\alpha}^{(t+1)}$, given by 
eq. (\ref{eq:alpha_estimate}), is independent of 
$\gamma_1, \ldots, \gamma_K$
and can be expressed in terms of $\bz$ and $\bG$.
Moreover, $\bB^{1,(t+1)}_h$, given by eq. (\ref{eq:B_under_H1}), 
is independent of 
$\gamma_1, \ldots, \gamma_K$
and can be expressed in terms of $\bz$ and $\bG$.
Thus
$\widehat{\gamma}_k^{1,(t+1)}$, given by
eq. (\ref{eq:gamma_h_estimate}),
can be expressed as
$\widehat{\gamma}_k^{1,(t+1)}=\gamma_k g^{1,(t+1)}_k(\bz, \bG)$
and the lemma is proved.

\section*{Proof of Theorem 2}

Iterative application of Lemma 2, starting with the assumption that 
$\widehat{\gamma}_k^{1,(0)}=\gamma_k g^{1,(0)}_k(\bz, \bG)$ and $\widehat{\gamma}_k^{0,(0)}=\gamma_k g^{0,(0)}_k(\bz, \bG)$, $k=1, \ldots, K$, 
is independent of $\gamma_1, \ldots, \gamma_K$, leads
to 
$\widehat{\gamma}_k^{1,(t_{\max})}=\gamma_k g^{1,(t_{\max})}_k(\bz, \bG)$ and $\widehat{\gamma}_k^{0,(t_{\max})}=\gamma_k g^{0,(t_{\max})}_k(\bz, \bG)$, $k=1, \ldots, K$. Then, it is sufficient to observe that
$\widehat{\alpha}^{(t_{\max})}$, given by eq. (\ref{eq:alpha_estimate_whitened}), is independent 
of $\gamma_1, \ldots, \gamma_K$
and can be expressed in terms of $\bz$,
$g_1^{1,(t_{\max}-1)}(\bz, \bG), \ldots, g_K^{1,(t_{\max}-1)}(\bz, \bG)$.
We can conclude that
the statistic of the proposed detector, given by eq. (\ref{eq:GLRT_approximation}),
is independent 
of $\gamma_1, \ldots, \gamma_K$ (under $H_0$).

\bibliography{group_bib}

\end{document}